\documentstyle[12pt]{article}
\begin{document}
\begin{center}
{\bf
SUM RULES FOR TOTAL INTERACTION CROSS SECTIONS OF RELATIVISTIC
ELEMENTARY ATOMS WITH ATOMS OF MATTER UP TO TERMS OF ORDER
{\large $\alpha^2$}\\}
\vspace*{1cm}
L.AFANASYEV, A.TARASOV and O.VOSKRESENSKAYA\\
{\it Joint Institute for Nuclear Research, \\
141980 Dubna,
Moscow Region,\\
Russia}
\end{center}

\vspace*{.5cm}
\begin{abstract}
{\small It is shown that $\alpha^2$-term of sum rules for the total cross
sections of interaction of elementary atoms with matter ones, obtained in
ref.\cite{C-2} is wrong. New sum rules valid up to terms of order $\alpha^2$
are derived.}
\end{abstract}

\vspace*{1cm}

The most full qualitative analysis of the interaction of relativistic
elementary atoms (EA) with matter in the Born approximation have been done
by S. Mrowczynski in the ref.\cite{C-1} In the ref.\cite{C-2} the sum rules
for the total cross sections of interaction of EA with target
(matter) atoms (TA) have being derived from general results of ref.\cite{C-1}
under some assumption, validity of which will be discussed below.

Since the all elastic scattering amplitudes, calculated in the Born
approximation, are real, one can't apply the optical theorem for the
calculation of the corresponding total cross sections.  Due to this reason
in the Born approximation the total cross sections calculated by summing of
all partial transition total sections $\sigma_{if}$

\begin{equation}
\sigma^{tot}(i)=\sum\limits_{f}\sigma_{fi}\,.
\end{equation}

In the considering case $\sigma_{if}$ being the cross sections of
transitions between some initial (i) and final (f) states of EA in the
screened Coulomb field of TA, described by potential $U(r)$. According to the
results of ref.\cite{C-1} they are presented in the form
\begin{equation}
\sigma_{fi}=\int \left|A_{fi}(\vec q)\right|^2d^2q_{\bot}\, ,
\end{equation}
where
\begin{equation}
A_{fi}(\vec q)=U(Q)[\rho_{fi}(\vec q)-\vec \beta \vec \gamma_{fi}(\vec q)]\,,
\end{equation}
\begin{equation}
U(Q)=4\pi\int U(r)\frac{\sin Qr}{Q}rdr\,,
\end{equation}
\begin{equation}
Q=\sqrt{{\vec q}^{\,2}-q_{0}^{2}}=\sqrt{{\vec
q}^{\,2}_{\bot}+q_{L}^{2}-q_{0}^{2}}= \sqrt{{\vec
q}^{\,2}_{\bot}+q_{L}^{2}(1-\beta^2)}\,,
\end{equation}
$$q_0=\varepsilon_f-\varepsilon_i+\frac{Q^2}{2M}=\beta q_L\,.$$
Here $\vec q_{\bot}$ and $q_L$ are the transverse and longitudinal
components of the vector $\vec q$ --- three dimension momentum transverse
to the target, $M$ is mass of EA,  $\beta$ is it's velocity in the lab
system.

The transition density $\rho_{fi}\,(\vec q)$ and transition current
$\vec j_{fi}\,(\vec q)$ are expressed in terms of the wave functions $\psi_i$
and $\psi_f$ of initial state of EA with help of relations:
\begin{equation}
\rho_{fi}(\vec q)=\int \rho_{fi}(\vec r)\left(
e^{i\vec q_{1}\vec r}-e^{-i\vec q_{2}\vec r}\right)d^3r,
\end{equation}
\begin{equation}
\vec j_{fi}(\vec q)=\int \vec j_{fi}(\vec r)\left(
\frac{\mu}{m_1}e^{i\vec q_{1}\vec r}+\frac{\mu}{m_2}e^{-i\vec q_{2}\vec
r}\right)d^3r,
\end{equation}
\begin{equation}
\rho_{fi}(\vec r)=\psi_{f}^{\ast}(\vec r)\psi_{i}(\vec r)\,,
\end{equation}
\begin{equation}
\vec j_{fi}(\vec r)=\frac{i}{2\mu}\left[ \psi_{i}(\vec
r)\vec \nabla\psi_{f}^{\ast}(\vec r)-\psi_{f}^{\ast}(\vec r)\vec \nabla
\psi_{i}(\vec r)\right],
\end{equation}
$$\vec q_1=\frac{\mu}{m_1}\vec q,\quad\vec q_2=\frac{\mu}{m_2}\vec q,\quad
\mu=\frac{m_1m_2}{M},\quad M=m_1+m_2; $$
where $m_{1,2}$ are the masses of elementary components of EA.

Since $q_L\approx\varepsilon_f-\varepsilon_i\sim
\mu\alpha^2\sim\alpha<q_{\bot}>$, where $<q_{\bot}>\,\sim\mu\alpha$ is
typical value of $q_{\bot}$ in the problem under consideration, on the first
sight it is natural to neglect the $q_L$-dependence of the transition
amplitude and to put
\begin{equation} A_{fi}(\vec q)\approx A_{fi}(\vec
q_{\bot})\,.
\end{equation}

Just this approximation have been made by authors\cite{C-2} in their
derivation of the following sum rules for the total EA --- TA cross sections:
\begin{equation}
\sigma^{tot}=\sigma^{elect}+\sigma^{magnet}\,,
\end{equation}
\begin{equation}
\sigma^{elect}=2\int U^2(q_{\bot})[1-S(q_{\bot})]d^2q_{\bot}\, ,
\end{equation}
\begin{equation}
S(q_{\bot})=\int |\psi(\vec r)|^2 e^{i\vec q_{\bot}\vec r}d^3r\,;
\end{equation}
\begin{equation}
\sigma^{magnet}=\int U^2(q_{\bot})K(q_{\bot})d^2q_{\bot}\, ,
\end{equation}
\begin{equation}
K(\vec q_{\bot})=\int \left (\frac{1}{m_{1}^{2}}+\frac{1}{m_{2}^{2}}+
\frac{2}{m_{1}m_{2}}e^{i\vec q\vec r}\right)
\left|\vec \beta~\vec \nabla~\psi_i(\vec r)\right|^2d^3r.
\end{equation}

Putting for the sake of qualitative estimations
$$U(r)=\frac{Z\alpha}{r}e^{-\lambda r},\quad \lambda\sim m_e\alpha
Z^{1/3},\quad \alpha=\frac{1}{137}\,;$$
it is easily to derive from (12 -- 15)
\begin{equation}
\sigma^{elect}\sim
\frac{Z^2}{\mu^2}\ln\biggl(\frac{\mu}{Z^{1/3}m_e}\biggl)\, ,
\end{equation}
\begin{equation}
\sigma^{magnet}\sim
\frac{Z^{4/3}\alpha^2}{m_e^2}\, ;
\end{equation}
where $m_e$ is electron mass.

Thus, although $\sigma^{magnet}$ is proportional to $\alpha^2$, squared
electron mass in the denominator of the eq.\ (17) make the relative
contribution of "magnetic" term not negligible small compared to electric
one, especially for the case of EA, composed from the heavy hadrons.
This result seems physically unreasonable, that forces us to reanalysis the
problem.

If is easily to verify that the subsequent members of expansion of
transition density and current over powers of small quantity $q_L$
\begin{equation}
\rho_{fi}=\sum_{n=0}^{\infty}\rho_{fi}^{(n)}\,,\quad
\rho_{fi}^{(n)}=\frac{q_{L}^{n}}{n!}\biggl(\frac{d^n}{dq_{L}^{n}}\rho_{fi}
\biggl)\biggl\vert_{q_L=0}\,,
\end{equation}
\begin{equation}
\vec j_{fi}=\sum_{n=0}^{\infty}\vec j_{fi}^{(n)}\,,\quad
\vec j_{fi}^{(n)}=\frac{q_{L}^{n}}{n!}\biggl(\frac{d^n}{dq_{L}^{n}}\vec
j_{fi} \biggl)\biggl\vert_{q_L=0}\,;
\end{equation} obey the following relations:
\begin{equation} \rho_{fi}^{(n)}\sim\alpha^n\,,\quad
j_{fi}^{(n)}\sim\alpha^{n+1}\,.
\end{equation}
Thus, neglecting the $q_L$-dependence of transition amplitude, the
authors\cite{C-2} have taking into
account only one term of order $\alpha$, namely $\vec \beta\cdot\vec
j_{fi}\,^{(0)}$ and have neglected another one $\rho_{fi}^{(1)}$.

The detail analysis of the problem shows that the contribution of terms
$\rho_{fi}^{(n+1)}$ and $\vec \beta\cdot\vec j_{fi}\,^{(n)}$ to the total
cross section are strongly destructively interfere, that results in
negligible contribution to it of all terms besides of $\rho_{fi}^{(0)}$.
With help of simple but enough cumbersome calculations one can derive the
following sum rules:
\begin{equation} \sigma^{tot}=2\int
U(q_{\bot})^2\left[1-S(q_{\bot})\right]d^2q_{\bot}- \int
U(q_{\bot})^2W(q_{\bot})d^2q_{\bot}+O(\alpha^4)\,,
\end{equation}
$$W(q_{\bot})=\frac{1}{M^2}\int \left(\vec \beta \vec
r\right)^2[q^4\psi_i^2(\vec r)+ \left(2\vec q_{\bot}\vec
\nabla\psi_i(\vec r)\right)^2]e^{i\vec q_{\bot}\vec r}d^3r\,.$$
Only the first term in the relation (21) is numerically significant.  The
second one, being of order $\alpha^2$, don't contain any enhancement factor
and may be neglected in the practical applications.  This justifies the
usage of the simple relation
\begin{equation} \sigma^{tot}=2\int
U(q_{\bot})^2\left[1-S(q_{\bot})\right]d^2q_{\bot}
\end{equation}
in the paper of authors\cite{C-3}.

This work is partially support by RFBR grant 97--02--17612.

\end{document}